\documentclass[conference]{IEEEtran}
\usepackage{amsmath}
\usepackage{amsfonts}
\usepackage{amssymb}
\usepackage{graphicx}
\usepackage{subcaption}
\usepackage{url}
\usepackage{calrsfs}
\newcommand{\mytilde}{\raise.17ex\hbox{$\scriptstyle\mathtt{‌​\sim}$}}

\begin{document}

\title{Performance Analysis of Physical Layer Network Coding for Two-way Relaying over Non-regenerative Communication Satellites}

\author{\IEEEauthorblockN{Saket D. Buch}
	\IEEEauthorblockA{Department of Electrical Engineering\\Indian Institute of Science\\
		Bangalore, India 560012\\
		Email: saket@ee.iisc.ernet.in}
	\and
	\IEEEauthorblockN{B. Sundar Rajan}
	\IEEEauthorblockA{Dept. of Electrical Communication Engineering\\Indian Institute of Science\\
		Bangalore, India 560012\\
		Email: bsrajan@ece.iisc.ernet.in}}

\maketitle

\begin{abstract}
Two-way relaying is one of the major applications of broadband communication satellites, for which an efficient technique is Physical Layer Network Coding (PLNC). Earlier studies have considered satellites employing PLNC with onboard processing. This paper investigates the performance of PLNC over non-regenerative satellites, as a majority of the operational and planned satellites do not have onboard processing. Assuming that the channel magnitudes of the two users are equal, two operating conditions are considered with uncoded-QPSK relaying. In the first condition, both users are completely synchronized in phase and transmit power, and in the second condition, phase is not synchronized. The peak power constraint imposed by the satellite amplifier is considered and the error performance bounds are derived for both the conditions. The simulation results for end-to-end Bit Error Rate (BER) and throughput are provided. These results shall enable communication system designers to decide system parameters like power and linearity, and perform trade-off analysis between different relaying schemes. 
\end{abstract}
\IEEEpeerreviewmaketitle
\section{Introduction}
Satellite communication is an attractive solution for extending the reach of broadband communications and cellular backhauling to rural and low population density areas \cite{minoli2015innovations}. To cater to increasing user demands, satellites are required to provide higher throughput within a limited bandwidth. An alternative explored to reduce bandwidth consumption is Network Coding \cite{vieira2010feasible}. A recent experiment demonstrated Network Coding in a video conferencing situation over a non-regenerative satellite \cite{dlrnetcod2010}. Further reduction in bandwidth is possible by using PLNC \cite{zhang2006hot}. This technique has been studied extensively for fading channels encountered in terrestrial communication. 

As the physical layer in a non-regenerative satellite communication system is different from terrestrial networks, several aspects need reconsideration. One such important difference is that of peak-power constraint during transmission from satellite to ground. Another difference is that unlike terrestrial communication, certain satellite channels are slow varying, which makes it possible to use precoding at users \cite{vazquez2016precoding}. For example, in \cite{rossetto2010comparison,vieira2010feasible}, PLNC is suggested only for advanced processing satellites. This excludes a large number of planned and operational communication satellites which are non-regenerative. In \cite{abuhaselperformance}, the performance of Analog Network Coding (ANC) over a non-regenerative satellite with nonlinear amplifiers is studied, but PLNC is not considered. To the best of the authors' knowledge there is no reference dealing with PLNC for non-regenerative satellites. 

This paper investigates the performance of PLNC over non-regenerative satellites. The contributions of this paper are:
\begin{itemize} 
	\item Performance bounds are provided for PLNC with and without phase synchronization amongst users in AWGN satellite channel. 
	\item Peak power constraint is considered at the satellite amplifier and its impact on transmission of superposed constellation to hub is investigated. 
	\item Degradation in Broadcast (BC) phase performance due to peak power constraint is investigated. This is also crucial in satellites using onboard processing.
\end{itemize}
The paper is organized as follows: The system overview and signal model are described in Section II. The impact of peak power constraint in the BC phase has been described in Section III. The bounds on performance of PLNC with and without phase synchronization have been described in Sections IV and V. The results of simulations for end-to-end throughput and Bit Error Rate are provided in Section VI. Section VII summarizes the inferences from the simulation and provides a list of topics for further research.
\section{System Overview and Signal Model}
Non-regenerative communication satellites are microwave repeaters which translate uplink frequency to downlink frequency and amplify the signal before relaying it to ground stations. If small terminals are used on both ground stations, the cascading of near-identical non-regenerating links \cite{hasna2003outage} leads to a 3 dB loss in effective SNR. Therefore, the satellite first transmits the signal from the user to a large ground station (hub) for regeneration before transmitting to the other user. The hub-satellite link is designed such that there is negligible degradation in user-satellite link SNR. That is, the weak user-satellite link is cascaded with a strong hub-satellite link \cite{maral2011satellite}. Thus the satellite network has a star network topology with hubs (or gateways) coordinating communication between users. The disadvantage is that the additional channel uses required for satellite-hub and hub-satellite links result in reduction of spectral efficiency by a factor of two. In this paper only single-beam satellites, or equivalently, users in a single beam of a multi-beam satellite are considered. Also, users, satellite, and the hub use only one antenna each to transmit and receive the signals.
\begin{figure}[h]
\centering
\includegraphics[width=21pc]{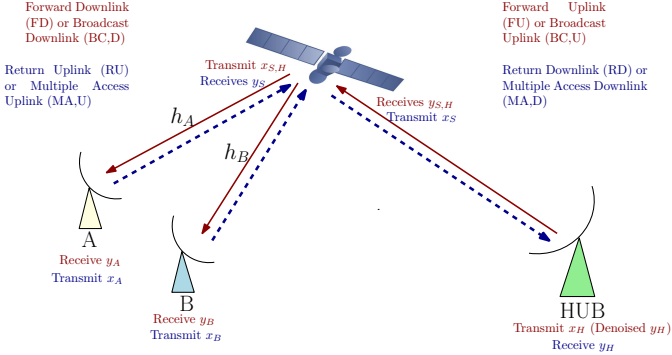}
\caption{Satellite Communication Links}
\label{SAT_LNK}
\end{figure}
\subsection{Satellite Links}
A link from user to hub is called \emph{Return link} and a link from hub to user is called \emph{Forward link}. Similarly, a link from ground station (hub or user) to satellite is called \emph{Uplink} and a link from satellite to ground station is called \emph{Downlink}. There are four links in a satellite network with star topology \cite{maral2011satellite}. The user to satellite link is called Return Uplink (\textit{RU}) and the satellite to hub link is called Return Downlink (\textit{RD}). The link from hub to satellite is called Forward Uplink (\textit{FU}) and from satellite to user is called Forward Downlink (\textit{FD}). In the PLNC case, since both users transmit together, the user to hub links are called Multiple Access (MA) links i.e. MA-Uplink and MA-Downlink. Similarly, the hub transmits a symbol which is common to both users and hence the hub to user links are called Broadcast (BC) links i.e. BC-Uplink and BC-Downlink. The links are shown in Fig. \ref{SAT_LNK}
\subsection{Signal Model}
As described earlier, PLNC relaying consists of two phases: Multiple Access (MA) and Broadcast (BC). The signal model for both phases is provided assuming users using QPSK signal sets. However, it is applicable to other modulations also. 
\subsubsection{Multiple Access Phase}
Consider users A and B wanting to exchange data through a satellite link. We shall follow the notation similar to \cite{shukla2012wireless}. Assume that A wants to send a 2-bit tuple to B and vice versa. The first phase of communication involves the links from the users to the hub. The users transmit complex symbols from constellation $\mathcal{S}$ which is a unit-energy QPSK (symmetric 4-PSK) constellation $ \{ \pm \frac{1}{\sqrt{2}} \pm \frac{j}{\sqrt{2}} \}$. Let  $\mu:\mathbb{F}_2^2 \rightarrow \mathcal{S}$ be the function mapping the bits to complex symbols. If $s_A, s_B \in \mathbb{F}_2^2$, then the transmitted symbols are $x_A = \mu(s_A), x_B = \mu(s_B) \in \mathcal{S}$. The constellation is labelled as per Gray labelling. We also define a function  $\lambda:\mathbb{F}_2^2 \rightarrow \mathbb{Z}_4$ mapping the bits to the set $\mathbb{Z}_4 = \{0,1,2,3\}$.  If $s_A, s_B \in \mathbb{F}_2^2$, then the symbol labels $m_A = \lambda(s_A), m_B = \lambda(s_B) \in \mathbb{Z}_4$.

Let the channel coefficients in the user-satellite links be $h_A$ and $h_B$ for user-A to satellite and user-B to satellite links respectively. We consider the ubiquitous Fixed Satellite Service (FSS) scenario, where, the user terminals are fixed and the line-of-sight channel is slow varying \cite{minoli2015innovations}. As channel amplitudes in such channels change at time-scales much longer than frame duration, power control can be implemented \cite{vazquez2016precoding}. Hence, perfect power control amongst user terminals has been assumed ($|h_A| = |h_B|$). Since both attenuation and amplification in the links are scaling factors, they are assumed to be unity. In addition to this, two cases are considered. In the first case, also called \emph{PLNC with precoding}, the users are phase synchronized. Thus, without loss of generality, $h_A = h_B = 1$. In the other case, phase is not synchronized. Hence, $h_B = h_Ae^{j\theta}$, where $\theta \mytilde \mathit{Unif}\left[0,2\pi\right)$. The satellite receives $y_S$, a noisy and scaled superposition of $x_A$ and $x_B$, given by
\begin{align}
y_S &= h_Ax_A + h_Bx_B + n_S,
\end{align}
where the additive noise $n_S$ is $\mathcal{CN}(0,\sigma_{MA,U}^2)$, and $\mathcal{CN}(0,\sigma^2)$ denotes the circularly symmetric complex Gaussian random variable with variance $\sigma^2$. The received signal is amplified by the High Power Amplifier (HPA) of the satellite during which the instantaneous magnitude of the signal is limited by the peak power constraint \cite{maral2011satellite}. However, for the perfect synchronization case the superposed constellation has four times the peak power of each user signal (see Appendix A). This amplifier saturation problem was pointed out in the context of Analog PNC in \cite{vieira2010feasible}. The amplified signal $x_S$ is transmitted from satellite and received at the hub as,
\begin{equation}
y_H = x_S + n_H = f(h_Ax_A + h_Bx_B + n_S,T) + n_H,
\end{equation}
where the additive noise $n_H$ is $\mathcal{CN}(0,\sigma_{MA,D}^2)$, and the function $f(.,T)$ is the transfer function of the non-linear amplifier with peak output magnitude constrained to $T$. It should be noted that hub is much larger than user terminals and hence the SNR of satellite-hub links is almost $10~dB$ higher than user-satellite links. For a given channel phase-shift $\theta$, the constellation received at the hub, $\mathcal{S}_r(\theta)$ is given as
\begin{equation}
\mathcal{S}_r(\theta) = \lbrace s_i + e^{j\theta}s_j | s_i,s_j \in \mathcal{S} \rbrace,
\end{equation}
where, $\theta$ is 0 for perfect synchronization. On this, the hub performs Maximum Likelihood (ML) decoding to estimate the transmitted pair $(x_A,x_B)$.
\begin{figure}[htbp]
	\centering
	\includegraphics[width=21pc]{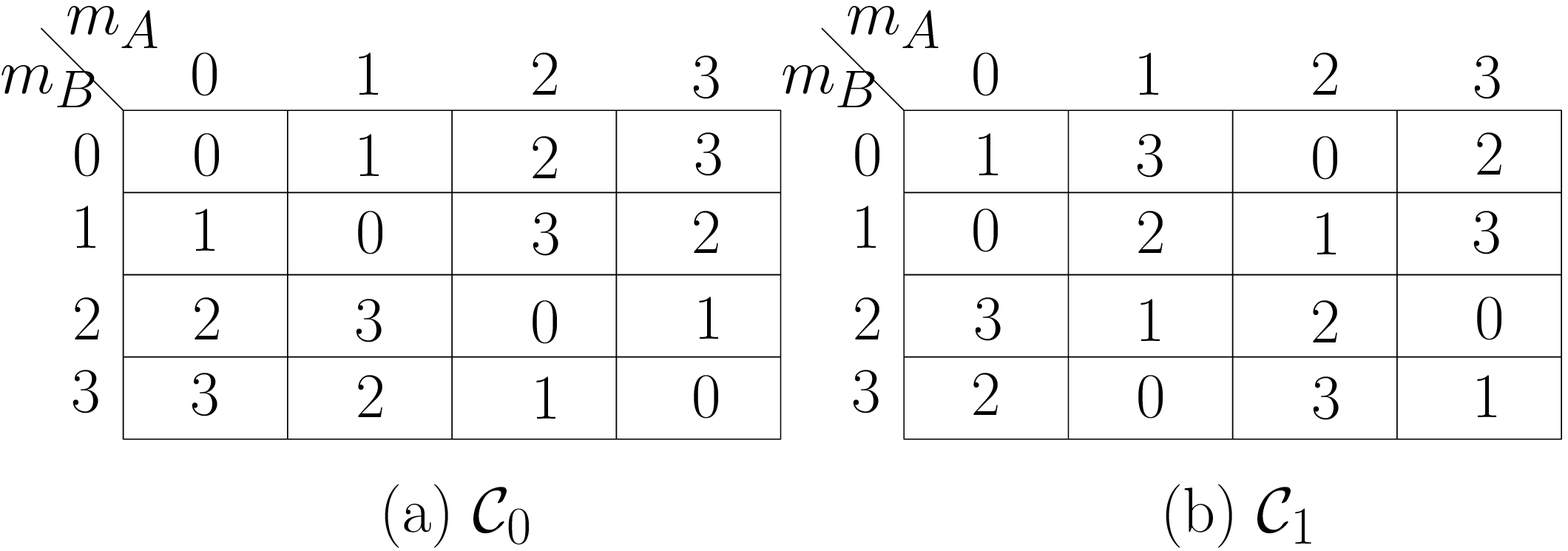}
	\caption{Denoising Maps for AWGN channel}
	\label{Den_Map}
\end{figure}
\subsubsection{Broadcast Phase}
Instead of transmitting the estimated pair, the hub applies a many-to-one map, also called a denoising map on the received constellation \cite{muralidharan2013wireless}. That is, the higher cardinality received constellation is mapped via a many-to-one map to a constellation of smaller cardinality, say $\mathcal{S}_{BC,\theta}$. For perfect power control and random phase scenario, only two many-to-one maps (of cardinality 4) need to be used. The first one is the bitwise XOR and the other is a rotated version of the same \cite{koike2009optimized}. Both maps are shown in Fig. \ref{Den_Map}. The denoised symbol is subsequently transmitted to the satellite. The received signal at the satellite $y_{S,H}$ is given as, 
\begin{equation}
y_{S,H} = x_H + n_{S,H},
\end{equation}
where $x_H \in \mathcal{S}_{BC,\theta}$, and the additive noise $n_{S,H}$ is $\mathcal{CN}(0,\sigma_{BC,U}^2)$. The satellite amplifies this signal (with peak magnitude constraint) and broadcasts it to users A and B, which receive,
\begin{align}
y_A &= x_{S,H} + n_A = f(x_H + n_{S,H},T) + n_A ~~~\text{and} \nonumber\\
y_B &= x_{S,H} + n_B = f(x_H + n_{S,H},T) + n_B,
\end{align}
where $n_A$ and $n_B$ are additive noise $\mathcal{CN}(0,\sigma_{BC,D}^2)$. For both the cases considered in the paper, a 4-point constellation (QPSK) is used in the BC phase. Thus, there is no distortion due to the peak magnitude constraint in the BC phase. 
\subsubsection{Non-linearity Model}
In the signal model described earlier, non-linearity is introduced in the system through onboard power amplifiers. It is possible to use predistortion to create highly linear power amplifiers, but the peak power constraint remains as a fundamental limitation. This motivates the amplifier to be modelled as a memoryless device that imposes a peak-power constraint without phase distortion  \cite{kayhan2012constellation}. 

Let $y_S$ represent the input to the amplifier and $T$ denote the peak magnitude allowed at the output. Without loss of generality, the gain of the amplifier is assumed to be unity. The amplifier is also assumed to be memoryless, and hence, the output is not dependent on input at any other time. The output $x_S = f(y_S,T)$ can be described as:
\begin{align}
|x_S|& = min(|y_S|,T), \nonumber \\
arg(x_S)& = arg(y_S).
\end{align} 
Since the constellation $\mathcal{S}$ comprises of unit magnitude complex symbols, the threshold is assumed to be $1$. This ensures that in the case without PLNC, the output is same as the input. 
\begin{figure}[htbp]
	\centering
	\begin{subfigure}{.22\textwidth}
		\centering
		\includegraphics[width=10pc]{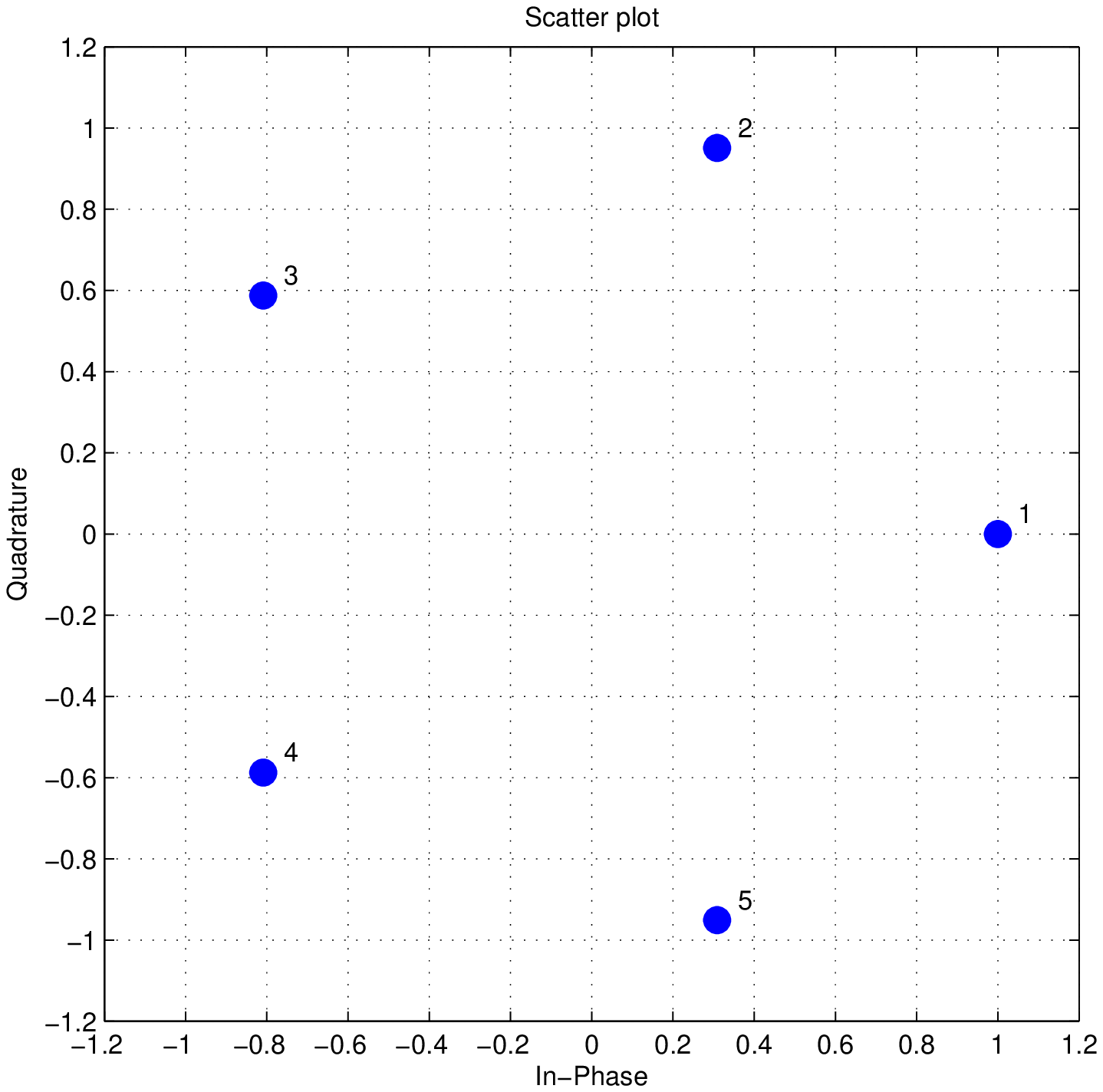}
		\caption{5-PSK}
		\label{5psk}
	\end{subfigure}%
	\begin{subfigure}{.22\textwidth}
		\centering
		\includegraphics[width=10pc]{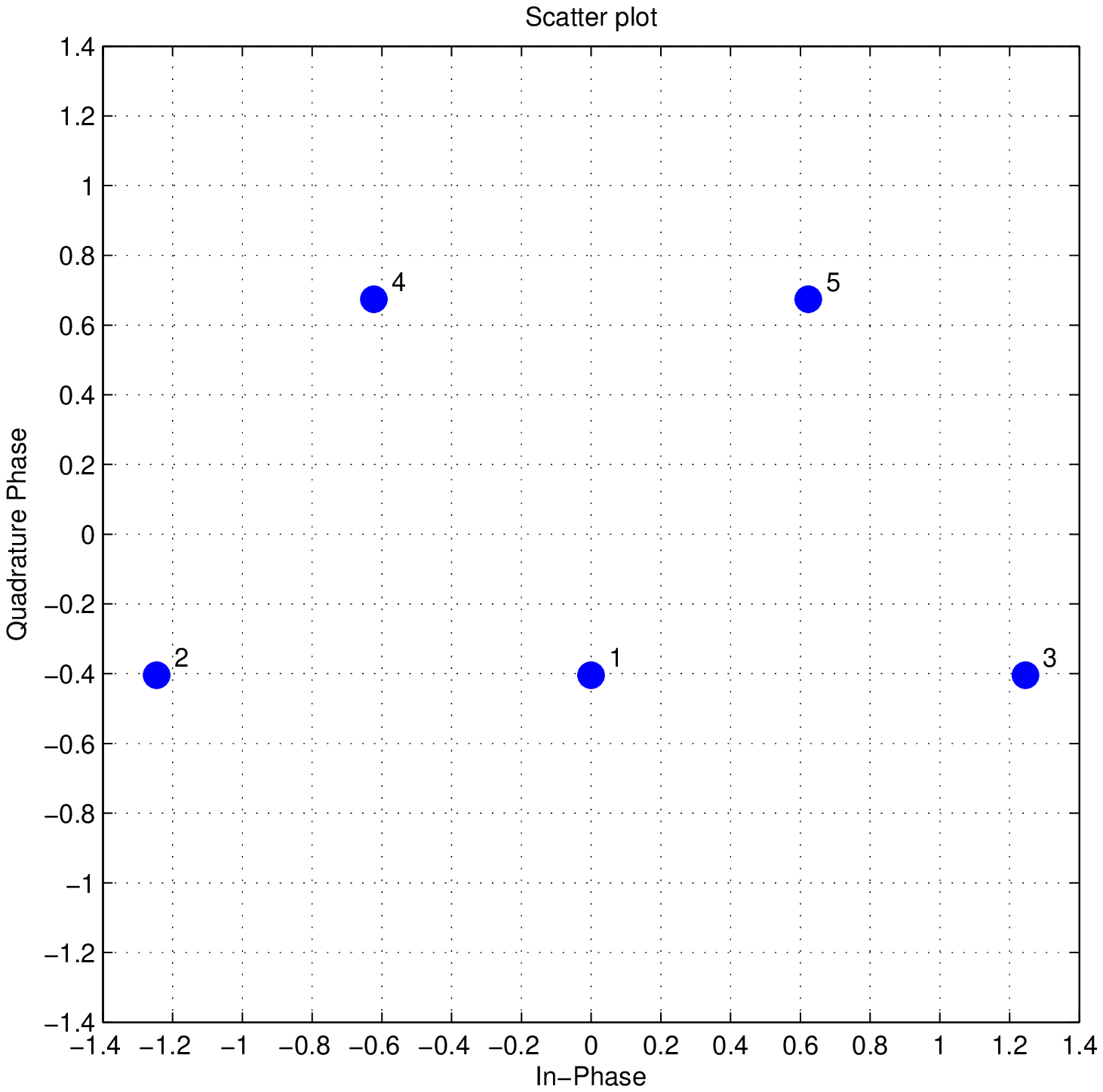}
		\caption{5-QAM}
		\label{5qam}
	\end{subfigure}
	\caption{Five point constellations}
	\centering
\end{figure}
\section{Broadcast Phase Constellations}
Two-way relaying with end-nodes using QPSK signal set requires the use of 5-point constellation in BC phase \cite{muralidharan2013wireless}. In general, depending on the number of distinct symbols required to complete the constrained partially-filled Latin squares for $M$-PSK modulations, the BC phase constellation may have \emph{non-standard} (i.e. not power of two) cardinality \cite{muralidharan2013wireless}. Such constellations, with an additional property of having good Euclidean distance can be obtained using the Greedy Sphere Packing algorithm \cite{koike2009optimized}. The algorithm maximizes the minimum Euclidean distance of a constellation under average power constraint and takes into account the probability of occurrence of each symbol in the denoising map. For QPSK relaying (with the probability distribution of symbols given in \cite{koike2009optimized}), we get the 5-QAM constellation shown in Fig \ref{5qam}. 

\begin{figure}[htbp]
	\centering
	\includegraphics[width=21pc]{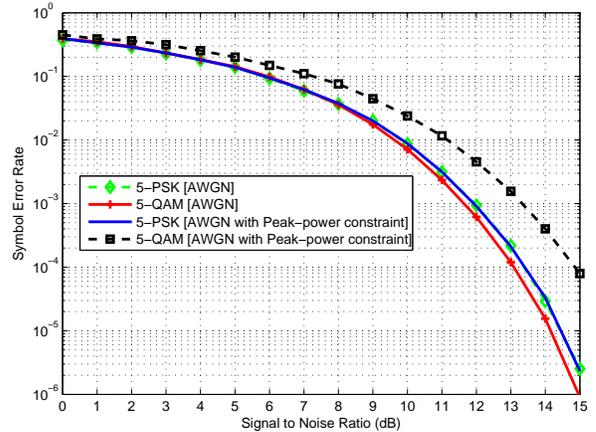}
	\caption{SER Performance of 5-PSK and 5-QAM constellations}
	\label{5QAM_Perf}
\end{figure}
The minimum distance of this 5-QAM constellation is higher than that of 5-PSK. This results in 0.5 dB gain in Symbol Error Rate (SER) performance over the 5-PSK. However, in a channel with peak power constraint, the higher Peak to Average Power Ratio (PAPR) of 5-QAM leads to poor performance compared to 5-PSK. The SER performance of both constellations with and without peak power constraint is shown in Fig. \ref{5QAM_Perf}. The input symbol distribution is the same as the one in \cite{koike2009optimized}. 

This sub-optimal performance of Greedy Sphere Packing can be attributed to its approach of building up the constellation, i.e. optimizing a subset of the constellation and then expanding it. It is clear that (using 5-QAM as an example) for a given constellation, the minimum distance of any of its subsets is always greater than or equal to the minimum distance of the complete constellation. The same is not true for PAPR. Thus, the constellation obtained in this manner is not guaranteed to be optimal in terms of PAPR. Since this paper only deals with AWGN channels with power control, QPSK is used in broadcast phase. However, if the power control is imperfect, the use of 5-point constellations may be necessary and 5-PSK should be used instead of 5-QAM since its performs better (though not proven to be optimum). Thus, the general problem of designing non-standard cardinality constellations with peak power constraint remains open.

\section{Performance of PLNC with Perfect Synchronization}
\begin{figure}[htbp]
\centering
\includegraphics[width=18pc]{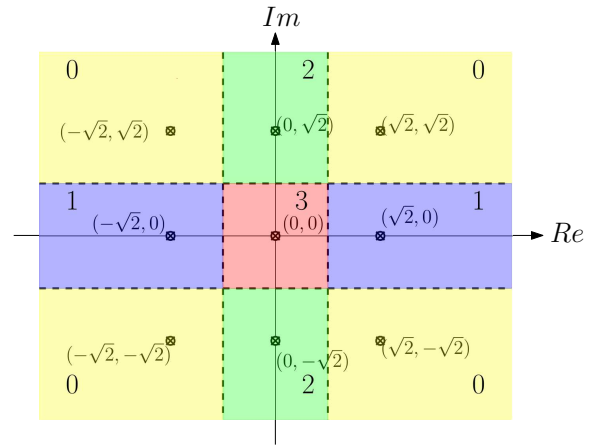}
\caption{Decision regions for 9-point constellation formed by superposition of two unit energy QPSK constellations with $h_A = h_B = 1$}
\label{Denoising}
\end{figure}

Perfect synchronization of users is equivalent to having Channel State Information at Transmitters (CSIT). This is not feasible with rapidly varying terrestrial communication channels. However, for fixed satellite terminals and geostationary satellites, the amplitude and phase vary slowly \cite{vazquez2016precoding}. This argument forms the justification for investigating this condition further. Under this assumption, the Bit-wise XOR function is the optimum denoising map for PSK signals \cite{koike2009optimized}. That is, the denoised symbol $x_H = \mu(s_H) \in \mathcal{C}$, where $s_H = s_A \bigoplus s_B$. The advantage here is that the minimum distance of the superposed constellation after applying the XOR map is same as that of the transmitted QPSK signal (see Fig.\ref{Denoising}). This implies that if the satellite performs many-to-one mapping onboard and transmits a symbol from QPSK constellation, or, the superposed constellation is sent to hub without impairments, the BER performance is same as the case where no PLNC is used, but the throughput is doubled \cite{zhang2006hot}. 

In onboard processing satellite, the signals are decoded at the satellite and depending on the channel coefficients, mapped to a symbol from a constellation of appropriate cardinality, which is transmitted. On the other hand, in a non-regenerative satellite, with star network topology, the superposed signals have to be transmitted to the hub for processing. The superposed constellation has higher PAPR as well as higher cardinality. In order to avoid clipping distortion from power amplifier, the superposed constellation is scaled. This leads to degradation in return downlink (satellite-to-hub) SNR. Considering these factors, the probability of bit error is evaluated based on the approach provided in \cite{prochazka2015relaying}. Let the bit error probabilities associated with the MA and BC phase be $P_{e,MA}$ and $P_{e,BC}$ respectively. A received bit is surely in error if there is an error in either MA or BC phase. Therefore, 
\begin{equation}
P_e \leq P_{e,MA} + P_{e,BC}.
\end{equation}
In the broadcast phase, two links with different SNRs are cascaded. As described in Section II, the reference scenario is a QPSK two-way relaying link without PLNC. First, a relation is established between SNR in the link with PLNC and SNR in the link without PLNC. Let the SNR of a given link be denoted by $\gamma$ with appropriate subscript. Since the BC link is the same as link without PLNC, 
\begin{eqnarray}
\gamma_{BC,U} = \gamma_{FU}, \nonumber \\
\gamma_{BC,D} = \gamma_{FD}.
\end{eqnarray}
The equivalent SNR of a link formed by cascading two non-regenerating links is given by \cite{hasna2003outage},
\begin{equation}
\gamma_{eq,BC} = \left[\left(1+\frac{1}{\gamma_{BC,U}}\right)\left(1+\frac{1}{\gamma_{BC,D}}\right) - 1\right]^{-1}.
\end{equation}
Therefore, the probability of bit error in the broadcast link is,
\begin{equation}
P_{e,BC} = Q\left(\sqrt{\gamma_{eq,BC}}\right).
\end{equation} 

In the MA link, the average energy of superposed constellation is twice that of transmitted QPSK constellation. Hence the received SNR at the satellite with PLNC is double compared to without PLNC. In the downlink, the signal is scaled in order to avoid clipping. This results in the average energy being reduced to half (see Appendix A). Thus, the relation between SNR in the links with and without PLNC are,
\begin{eqnarray}
\gamma_{MA,U} = 2 \times \gamma_{RU}, \nonumber \\
\gamma_{MA,D} = \frac{1}{2} \times \gamma_{RD}.
\end{eqnarray}
Scaling does not change the constellation, hence the equivalent SNR can be calculated as,
\begin{equation}
\gamma_{eq,MA} = \left[\left(1+\frac{1}{\gamma_{MA,U}}\right)\left(1+\frac{1}{\gamma_{MA,D}}\right) - 1\right]^{-1}.
\end{equation}
The final step is to evaluate the error performance of the 9-point received constellation. The decision regions of the received constellation along with the corresponding mapping (clustering) at the relay are shown in Fig. \ref{Denoising}. The map from integers ${0,1,2,3}$ indicated in the figure correspond to the two-bit tuples for $x_H$ as ${00,01,10,11}$ respectively.

Let $E$ be the average energy of the transmitted QPSK constellation. The points mapped to $0$ in the 9-point constellation, have two neighbors at a distance $d_{1} = \sqrt{2E}$, and one at a distance $d_{2} = 2\sqrt{E}$. Similarly, points mapped to $1$ and $2$ have three neighbors at a distance $d_{1}$ and two at a distance $d_2$. The central point mapped to $3$ has four neighbors at a distance  $d_{1}$ and four neighbors at a distance $d_2$. Let the superposed constellation be denoted as $\mathcal{S}_s$. If the probability of a point $x_i \in \mathcal{S}_s$ is denoted by $P_{x_i}$, the average probability of clustering error \cite{muralidharan2013wireless} is,
\begin{align}
P_{e,D} &= \sum_{i=1}^{9} P_{x_i} \sum_{x_j \neq x_i} P(x_i \rightarrow x_j)~~~~~~~x_i,x_j \in \mathcal{S}_s\\
 &= \frac{4}{16} \times \left(2Q\left( \sqrt{\frac{E}{N_0}}\right) + Q\left( \sqrt{\frac{2E}{N_0}}\right)\right) \nonumber\\&+ \frac{8}{16} \times \left(3Q\left( \sqrt{\frac{E}{N_0}}\right) + 2Q\left( \sqrt{\frac{2E}{N_0}}\right)\right) \nonumber\\&+ \frac{4}{16} \times \left(4Q\left( \sqrt{\frac{E}{N_0}}\right) + 4Q\left( \sqrt{\frac{2E}{N_0}}\right)\right)  \nonumber \\
 &=   3Q\left( \sqrt{\frac{E}{N_0}}\right) + \frac{9}{4}Q\left( \sqrt{\frac{2E}{N_0}}\right). 
\end{align}
A clustering error can result in multiple bits being in error. Considering the worst case, we take $P_{e,MA} = P_{e,D}$.
\begin{figure}[htbp]
	\centering
	\includegraphics[width=19pc]{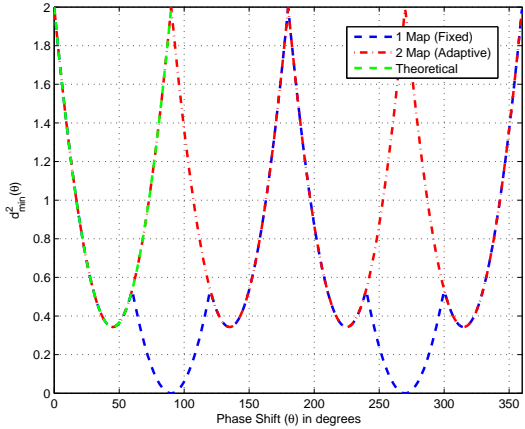}
	\caption{Minimum Squared Euclidean Distance as a Function of Channel Phase Shift in QPSK PLNC}
	\label{min_dist_awgn}
	\end{figure} 

\section{PLNC without Phase Synchronization}
In this section, the performance of QPSK PLNC is analyzed when precoding is not possible at the users. That is, the users are not phase synchronized. The minimum squared Euclidean distance is no longer $2E$, but a function of the channel phase shift $\theta$. The advantage of using two maps adaptively instead of one fixed map (in terms of minimum distance) is shown in Fig. \ref{min_dist_awgn}. It is obvious from this figure that this will result in poorer BER performance compared to the precoded case. The performance of broadcast link remains same as the precoded case. Thus, analysis is directed to evaluating the probability of error in the MA link. 
\begin{figure*}[htbp]
	\centering
	\begin{subfigure}{.5\textwidth}
		\centering
		\includegraphics[width=21pc]{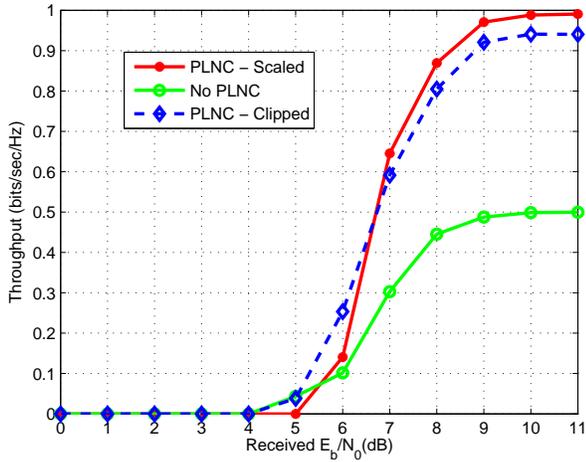}
		\caption{End-to-end Throughput}
		\label{fig5:sub1}
	\end{subfigure}%
	\begin{subfigure}{.5\textwidth}
		\centering
		\includegraphics[width=21pc]{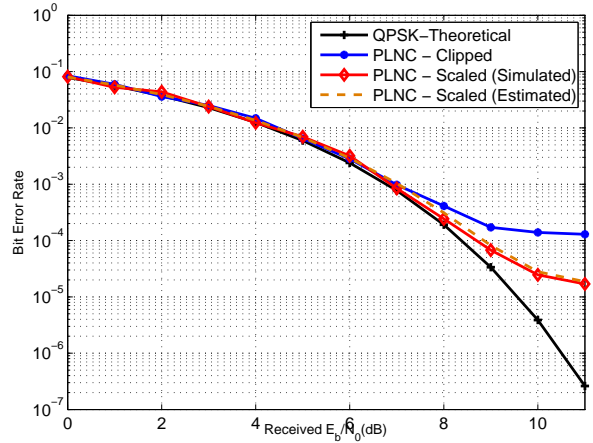}
		\caption{End-to-end BER}
		\label{fig5:sub2}
	\end{subfigure}
	\centering
	\caption{System performance assuming perfect synchronization and varying $\gamma_{FD}$}
	\label{fig5:test}
\end{figure*}
\begin{figure}[htbp]
	\centering
	\includegraphics[width=20pc]{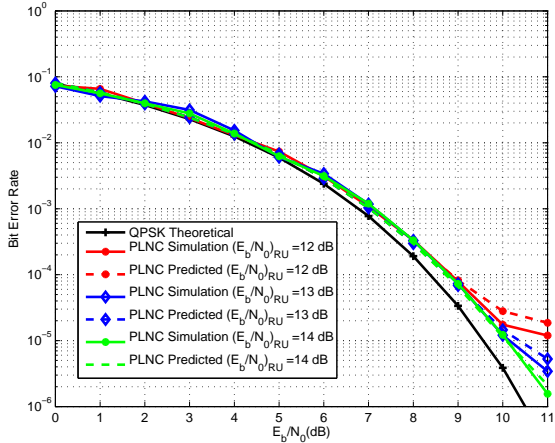}
	\caption{End-to-End BER performance for different $\gamma_{RU}$ and varying $\gamma_{FD}$}
	\label{BERRTN}
\end{figure} 
For a phase shift $\theta$, the received constellation at the relay is denoted by $\mathcal{S}_r(\theta)$. Consider points $x_k,x_j \in \mathcal{S}_r(\theta)$ such that they belong to different clusters. If the noise power is $N_0$, the signal-to-noise-ratio (SNR) is defined as $E/N_0$ and the probability of clustering error for a given $\theta$ is,
\begin{align}
P_{e,\theta}(x_k \rightarrow x_j) &= Q\left(\frac{|x_k-x_j|}{\sqrt{2N_0}}\right)\nonumber\\
& \leq Q\left(\frac{d_{min}(\theta)}{\sqrt{2N_0}}\right)\nonumber\\
& \leq \frac{1}{2}e^{-\frac{d_{min}^2(\theta)}{4N_0}}~~\text{since}~Q(x) \leq \frac{1}{2}e^{-\frac{x^2}{2}}.
\end{align}
Since the channel phase is random, with uniform probability distribution ($P_{\Theta}(\theta) = \frac{1}{2\pi}$; $0 \leq \theta < 2\pi$), the average probability of error for a given pair is,
\begin{align}
P_e(x_k \rightarrow x_j) &= \int_{0}^{2\pi} P_{e,\theta}(x_k \rightarrow x_j)P_{\Theta}(\theta) d\theta \nonumber \\
&= \frac{1}{4\pi}\int_{0}^{2\pi} e^{-\frac{d_{min}^2(\theta)}{4N_0}}d\theta. \nonumber
\end{align}
By symmetry (see Fig. \ref{min_dist_awgn}), and using Appendix B,
\begin{align}
P_e(x_k \rightarrow x_j) = \frac{1}{\pi}\int_{0}^{\frac{\pi}{2}} e^{-(1.5 - (cos\theta+sin\theta))SNR}d\theta,
\end{align}
where $SNR$ is calculated by cascading the MA uplink and downlink. It can be observed that there are $8$ points, each having two closest neighbors belonging to a different cluster. The probability of each point occurring is $1/16$ and hence,
\begin{equation}
P_{e,MA} = \frac{1}{\pi}\int_{0}^{\frac{\pi}{2}} e^{-(1.5 - (cos\theta+sin\theta))SNR}d\theta.
\end{equation} 
This integral is evaluated numerically at required SNR. The remaining analysis is same as that in the previous section and hence omitted. 
\section{Simulation Results}
For a given a transmission frame length, the throughput of the communication system is defined as a function of the Frame Error Rate (FER) and the spectral efficiency ($\eta$) \cite{muralidharan2013wireless} as,
\begin{equation}
Throughput = (1-FER) \times \eta~~~~bits/s/Hz.
\end{equation}
To make the simulations consistent with existing results \cite{muralidharan2013wireless,koike2009optimized} we consider a packet size of 256 symbols. The BER is calculated by considering at least 100 error instances at each step of $\frac{E_b}{N_0}$. Unless otherwise stated, the constellations are scaled before non-linearity to avoid clipping. Also, the simulations are for hub-based satellite system with onboard non-linearity. The parameters of the system without PLNC are as follows:
\begin{align}
\gamma_{RU} &= 15~dB~~~~\gamma_{RD} = 23~dB \nonumber\\
\gamma_{FU} &= 23~dB~~~~\gamma_{FD} = 3-15~dB
\end{align}

\begin{figure}[htbp]
	\centering
	\includegraphics[width=21pc]{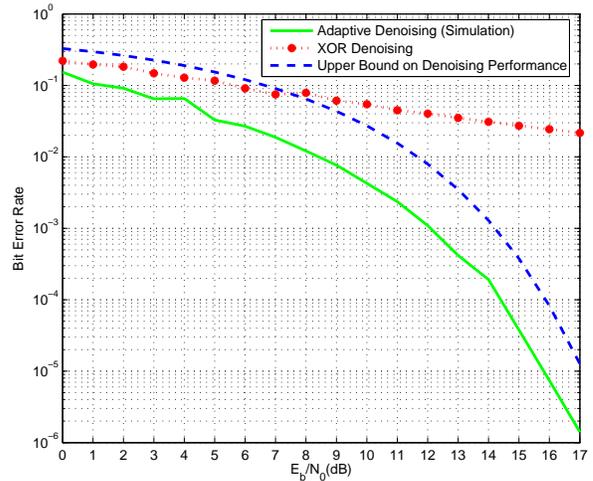}
	\caption{BER performance at relay for PLNC without phase synchronization and varying $\gamma_{RU}$}
	\label{PLNC_wo_PS}
\end{figure}  

\begin{figure}[htbp]
	\centering
	\includegraphics[width=21pc]{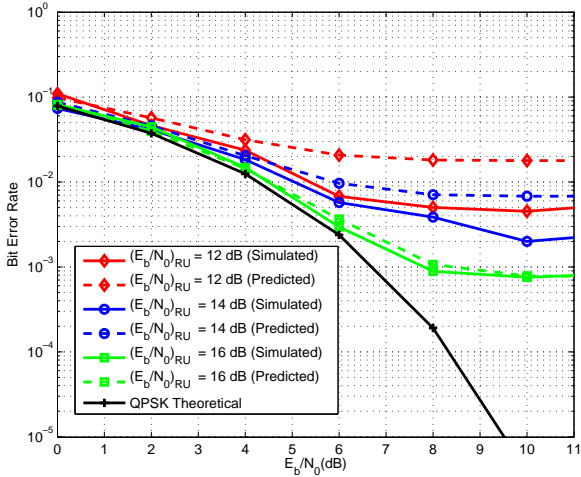}
	\caption{End-to-end BER performance of PLNC without phase synchronization while varying $\gamma_{FD}$ and different $\gamma_{RU}$}
	\label{PLNC_wo_PS_2}
\end{figure}
\subsection{QPSK PLNC with Phase Synchronization} 
The throughput performance for PLNC with perfect synchronization is provided in Fig. \ref{fig5:sub1} and the BER performance is provided in Fig. \ref{fig5:sub2}. It is seen that with two way relaying using QPSK and PNC a throughput of 1 bits/s/Hz is achieved using the same bandwidth as QPSK without PNC. It is clear from the BER curves that scaling is better than clipping. To compensate for the increased BER caused by peak power constraint a penalty has be to paid in terms of increased user uplink power. The estimated and simulated BER performances for different uplink power levels are shown in Fig. \ref{BERRTN}. It should be noted that the penalty is paid only in terms of uplink power. That is, the transmit power requirements at the satellite and hub are the same as the case when PLNC is not used.      
\subsection{QPSK PLNC without Phase Synchronization}
The BER performance at the relay is given in Fig. \ref{PLNC_wo_PS}. It can be seen that a fixed XOR map has very poor performance in this channel. For adaptive mapping, the map $\mathcal{C}_0$ is chosen if the channel phase shift $\theta$ is in the range $0~to~\pi/4$, $3\pi/4~to~5\pi/4$, and $7\pi/4~to~2\pi$. For all other $\theta$, $\mathcal{C}_1$ is used. Although this results in an improved performance, it is still poor compared to the perfect synchronization case. The end-to-end BER performance for PLNC without phase synchronization is provided in Fig. \ref{PLNC_wo_PS_2}. The upper bound also confirms the trend of the BER curve and becomes tighter as the user uplink $E_b/N_0$ is increased.

\begin{figure*}[!htbp]
\centering
\begin{subfigure}{.5\textwidth}
  \centering
  \includegraphics[width=15pc]{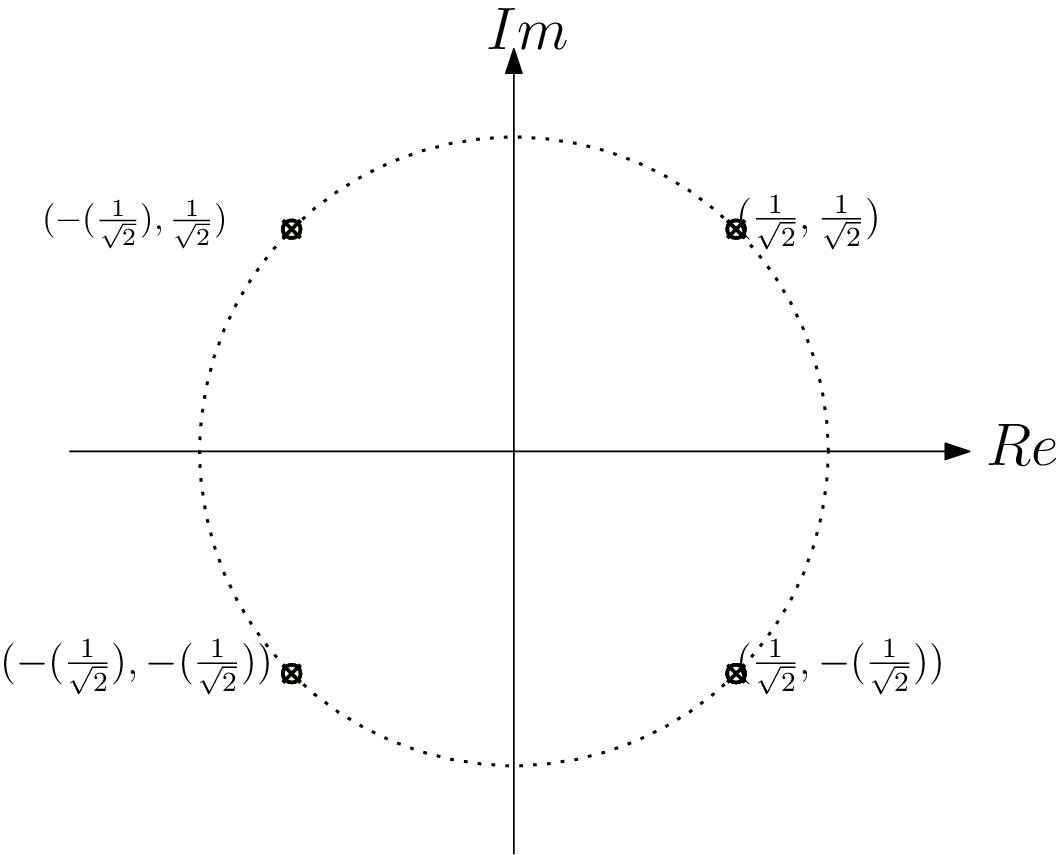}
  \caption{Unit Energy QPSK Constellation}
  \label{fig1:sub1}
\end{subfigure}%
\begin{subfigure}{.5\textwidth}
  \centering
  \includegraphics[width=15pc]{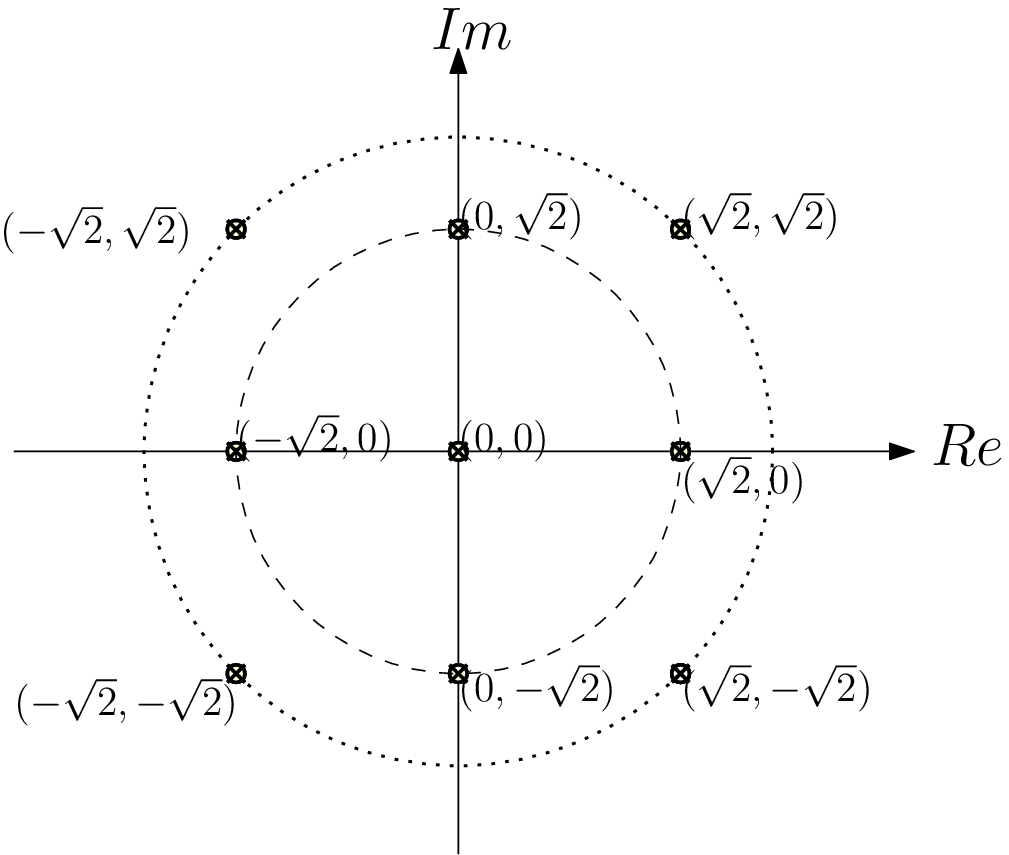}
  \caption{Superposed Constellation }
  \label{fig1:sub2}
\end{subfigure}
\centering
\caption{Constellations (without noise) before nonlinearity ($h_A = h_B = 1$)}
\label{fig1:test}
\end{figure*}

 \begin{figure*}[!htbp]
\centering
\begin{subfigure}{.5\textwidth}
  \centering
  \includegraphics[width=15pc]{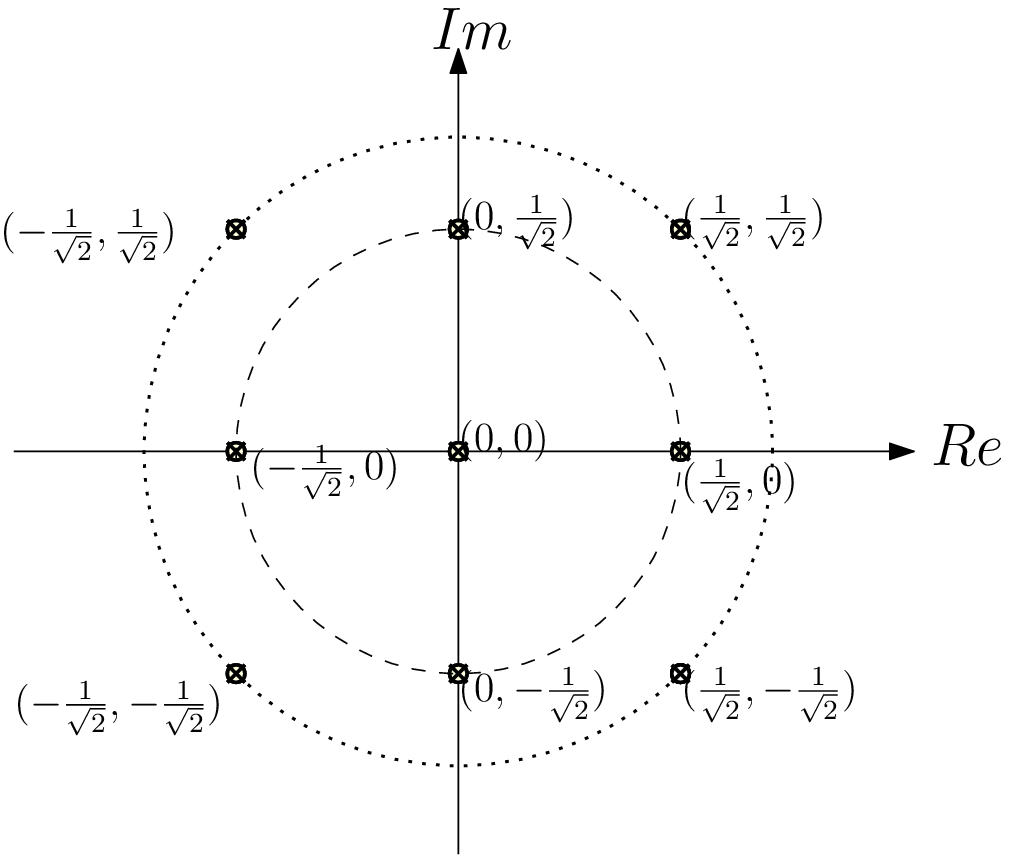}
  \caption{Scaled by a factor of 2 and then clipped}
  \label{fig2:sub1}
\end{subfigure}%
\begin{subfigure}{.5\textwidth}
  \centering
  \includegraphics[width=15pc]{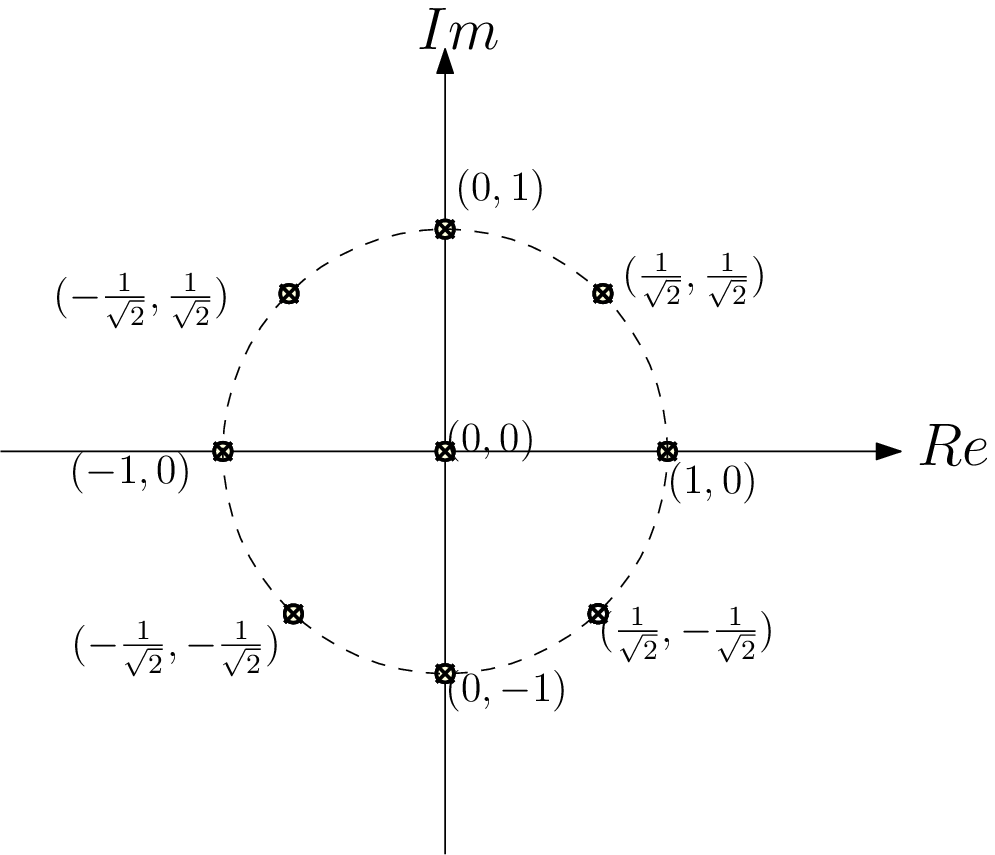}
  \caption{Clipped without scaling}
  \label{fig2:sub2}
\end{subfigure}
\centering
\caption{Superposed constellation after non-linearity}
\label{fig2:test}
\end{figure*} 

\section{Discussion and Future Work}
It is shown that PLNC with phase synchronization doubles the throughput for non-regenerative satellites inspite of the peak power constraint. Considering the bandwidth needed for hub links, end-to-end throughput is half that of the case when onboard processing is used. However, it is better than the case where PLNC is not used at all. In terms of handling the peak power constraint, clipping has poor performance when compared to scaling. When users are perfectly synchronized, the analysis of error performance of scaling scheme shows that the Return Uplink (Multiple Access Uplink) $\frac{E_b}{N_0}$ has to be increased to keep the BER at the same level as when PLNC is not used. When users are not phase synchronized, the penalty in BER is much higher. This provides a strong motivation to explore algorithms and techniques for precoding. The present work brings out several interesting problems that can be investigated:
\begin{itemize}
\item The analysis in this paper can be extended to other constellations. 
\item As mentioned in Section III, the problem of designing constellations with non-standard cardinality that are also optimal in the sense of PAPR needs to be investigated.
\item PLNC may need to be used for users in different beams of a multi-beam satellite. Due to the increasing number of multi-beam satellites \cite{minoli2015innovations}, this problem is very relevant. 
\item The present work considered users, hub, and satellite using only one antenna for transmit and receive. The concept of PLNC for satellite based MIMO can be explored.
\item Terrestrial wireless systems such as LTE also feature non-regenerating relay with peak power constraint \cite{iwamura2010relay} fow which the present work can be extended.
\end{itemize}

\appendices
\section{Scaling and Clipping of Superposed QPSK Constellations}
Let the peak energy $E_{peak}$ and average energy $E_{avg}$ of a constellation $\mathcal{S}$ be defined as 
\begin{align}
E_{peak} &=  \max_{x \in \mathcal{S}}  (|x|^2), \\
E_{avg} &= \sum_{x \in \mathcal{S}} p_x|x|^2,
\end{align}
where $p_x$ is the probability that the constellation point $x$ is transmitted. For a unit-energy QPSK signal, the peak energy and average energy are equal to $1$. In general, combination of two 4-point constellations can result in a constellation with a cardinality up to 16. Due to perfect synchronization of transmitters and no relative phase shift or scaling in channel, multiple pairs get mapped to the same point in the received constellation. The resulting 9-point constellation is shown in Fig. \ref{fig1:sub2}. This signal has peak energy $4$ and average energy $2$. The QPSK signal and superposed signal are shown in Fig. \ref{fig1:sub1} and \ref{fig1:sub2} respectively.  

However, the maximum magnitude of the complex symbol at the output of the non-linear device is restricted to $1$. To avoid clipping, the signal is scaled before the non-linear amplifier input. Since the peak magnitude is twice the threshold value, the signal magnitude is scaled by $2$ (\ref{fig2:sub1}). Then, the peak and average energy are,
\begin{align}
E_{peak}^{s}  &= 1, \nonumber\\
E_{avg}^{s} &= \frac{4}{16} \times 0 + \frac{8}{16} \times \frac{1}{2} + \frac{4}{16} \times 1 = \frac{1}{2}.
\end{align}
This implies that even though the amplifier is capable to providing average energy $1$, the signal is only able to extract half of it. This results in a $3~dB$ SNR reduction in that link compared to when PLNC is not used. 

If the signal is not scaled, the nonlinear device clips the input signal. In that case, (Fig. \ref{fig2:sub2}) the peak energy and average energy are,
\begin{align}
E_{peak}^{c} &= 1, \nonumber\\
E_{avg}^{c} &= \frac{12}{16} \times 1 + \frac{4}{16} \times 0 = 0.75.
\end{align}
The average energy in this case is greater than that of scaled signal, which generally indicates better performance. That is not the case here because noise variance is not scaled uniformly during the clipping of noisy signals. For example, noise around the constellation point $0+j0$ is accumulated over MA uplink and downlink without getting clipped, which degrades the SNR. On the other hand, in scaling, the noise around each point (from the first link) is also scaled along with the signal. Thus, SNR is not affected.
\begin{figure}[!htbp]
	\centering
	\includegraphics[width=18pc]{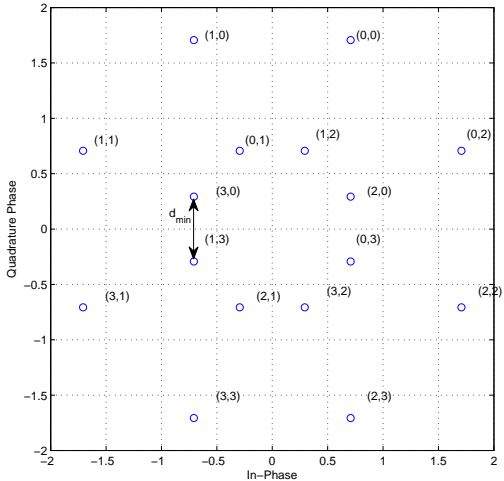}
	\caption{Received constellation at relay (hub), when $\theta = \pi/4$}
	\label{const_theta}
\end{figure}
\section{Minimum Distance for QPSK PLNC without Phase Synchronization}
The constellation seen at the relay, $\mathcal{S}_r(\theta)$, is formed from 16 pairs of transmit symbols. The many-to-one map at the relay maximizes the minimum of the distances between a pairs of points in $\mathcal{S}_r(\theta)$ which do not belong to the same cluster. Consider a pair of points in the received constellation, $(3,0)$ and $(1,3)$ shown in Fig. \ref{const_theta}. These pairs are chosen such that they belong to different clusters in both maps $\mathcal{C}_0$ and $\mathcal{C}_1$ (refer Fig. \ref{Den_Map}) and have minimum distance. With varying phase, the squared Euclidean distance between the pairs changes. Consider that $\theta$ varies from $0$ to $\pi/2$. In that range the minimum squared Euclidean distance, $d_{min}^2(\theta)$, is given as
\begin{align}
d_{min}^2(\theta) &= \frac{1}{2}((-1+j)-e^{j\theta}(1+j) \nonumber\\&+ (1+j) + e^{j\theta}(1+j))^2 \nonumber \\
 &= 2\left((-cos\theta+sin\theta) +j(1 - cos\theta-sin\theta)\right)^2 \nonumber\\
 &= 2\left((-cos\theta+sin\theta)^2 +(1 - cos\theta-sin\theta)^2 \right) \nonumber\\
 & = 2\left(3-2(cos\theta+sin\theta)\right).
\end{align}
As expected (see Fig. \ref{min_dist_awgn}), at $\theta=0$ and $\theta=\pi/2$, the squared minimum distance is $2$. At $\theta=\pi/4$, $d_{min}^2(\theta)$ attains its minimum value of \mytilde$0.34$.
\section*{Acknowledgment}
This work was supported partly by the Science and Engineering Research Board (SERB) of Department of Science and Technology (DST), Government of India, through J. C. Bose National Fellowship to B. Sundar Rajan.


\begin{thebibliography}{1}

\bibitem{minoli2015innovations}
D. Minoli, \emph{Innovations in Satellite Communication and Satellite
  Technology}. John Wiley \& Sons, 2015.

\bibitem{vieira2010feasible}
F. Vieira, S. Shintre, and J. Barros, ``How feasible is network coding in
  current satellite systems?'' in \emph{Proc. 5th ASMA/ 11th SPSC}, Sept. 2010, pp. 31--37.

\bibitem{dlrnetcod2010}
H. Bischl, H. Brandt, and F. Rossetto, ``An experimental demonstration of Network Coding for satellite networks,'' \emph{CEAS Space Journal}, vol. 2, no. 1-4, pp. 75--83, 2011.

\bibitem{zhang2006hot}
S. Zhang, S. C. Liew, and P. P. Lam, ``Hot topic: Physical-Layer Network Coding,'' in \emph{Proc. ACM Annu. Int. Conf. Mobile Comput. Network.,}, Los Angeles, CA, USA, 2006, pp. 358--365.

\bibitem{vazquez2016precoding}
M. {\'A}. V{\'a}zquez, et al., ``Precoding in
  multibeam satellite communications: Present and future challenges,''
  \emph{IEEE Wireless Comm.}, vol. 23, no. 6, pp. 88--95, 2016.

\bibitem{rossetto2010comparison}
F. Rossetto, ``A comparison of different physical layer network coding
  techniques for the satellite environment,'' in \emph{Proc. 5th ASMA/ 11th SPSC}, Sept. 2010, pp. 25--30.

\bibitem{abuhaselperformance}
K. A. Abuhasel, A. A. Khan, A. M. Hussein, and I. Ahmad, ``Performance of
  analog network coding based satellite systems in the presence of high power
  amplifier nonlinearities,'' in \emph{14th International Conference on
  Nonlinear Analysis, Nonlinear Systems and Chaos (NOLASC '15)}, 2015.

\bibitem{hasna2003outage}
M. O. Hasna and M.-S. Alouini, ``Outage probability of multihop transmission
  over Nakagami fading channels,'' \emph{IEEE Commun. Lett.}, vol. 7,
  no. 5, pp. 216--218, May 2003.

\bibitem{maral2011satellite}
G. Maral and M. Bousquet, \emph{Satellite communications systems: systems,
  techniques and technology}, John Wiley \& Sons, 2011.

\bibitem{shukla2012wireless}
S. Shukla, V. T. Muralidharan, and B. S. Rajan, ``Wireless network-coded
  three-way relaying using latin cubes,'' in \emph{Proc. of 23rd IEEE PIMRC}, Sydney, Australia, 2012, pp. 197--203.

\bibitem{muralidharan2013wireless}
V. T. Muralidharan, V. Namboodiri, and B. S. Rajan, ``Wireless Network-Coded Bidirectional Relaying Using Latin Squares for M-PSK Modulation,'' \emph{IEEE Trans. Inf. Theory}, vol. 59, no. 10, pp. 6683--6711, 2013.

\bibitem{koike2009optimized}
T. Koike-Akino, P. Popovski, and V. Tarokh, ``Optimized constellations for two-way wireless relaying with physical network coding,'' \emph{IEEE J. Select. Areas Commun.}, vol. 27, no. 5, pp. 773--787, 2009.

\bibitem{kayhan2012constellation}
F. Kayhan and G. Montorsi, ``Constellation design for transmission over
  nonlinear satellite channels,'' in \emph{Proc. of IEEE GLOBECOM}, 2012, pp. 3401--3406.

\bibitem{prochazka2015relaying}
P. Prochazka, T. Uricar, D. Halls, and J. Sykora, ``Relaying in butterfly
  networks: Superposition constellation design for wireless network coding,''
  in \emph{2015 IEEE International Conference on Communication Workshop
  (ICCW)}, London, 2015, pp. 2168--2174.

\bibitem{iwamura2010relay}
M. Iwamura, H. Takahashi, and S. Nagata, ``Relay technology in LTE-Advanced,''
  \emph{NTT DoCoMo Technical Journal}, vol. 12, no. 2, pp. 29--36, 2010.

\end{thebibliography}
\end{document}